\documentclass[article]{appolb}
\usepackage{graphicx}

\begin{document}
\title{{\WERS{Twist Expansion of\\ Forward Drell--Yan Process}}%
\thanks{Presented at EDS Blois 2015: The $16^{\mathrm{th}}$ Conference on Elastic and Diffractive\hfil\break
Scattering, Borgo, Corsica, France, June 29--July 4, 2015.}}
\headtitle{Twist Expansion of Forward Drell--Yan Process}
\author{{Tomasz Stebel},
{Leszek Motyka},
{Mariusz Sadzikowski}
\address{The Marian Smoluchowski Institute of Physics, Jagiellonian University\\
Łojasiewicza 11, 30-348 Kraków, Poland}}
\headauthor{T.~Stebel, L.~Motyka, M.~Sadzikowski}
\maketitle
\begin{abstract}

We present a twist expansion of differential cross sections of the forward Drell--Yan
process at the high energies. The expansion of all invariant form factors is performed
assuming Golec-Biernat and W{ü}sthof (GBW) saturation model and the saturation scale
plays the role of the hadronic scale of Operator Product Expansion (OPE). Some explicit
predictions for LHC experiments are given. It is shown also how the Lam--Tung relation is
broken at twist 4 what provides a sensitive probe for searching of higher twists.
\end{abstract}

\section{Introduction}

The Large Hadron Collider (LHC) opens new kinematic regions in high energy physics. The
most promising process at the LHC for investigating QCD effects for small Bjorken-$x$
($\simeq 10^{-6}$) and moderate energy scales is a forward Drell--Yan (DY) scattering. Such
a small $x$ at parton density scale $\mu^2 > 6$~GeV$^2$ is about two orders of magnitude
smaller than in measurements at HERA. Due to the forward kinematics of the process, the
LHCb detector is the most suitable for those measurements~\cite{lhcb0}.

The  angular distribution of DY lepton--antilepton pairs may be param-etrized in terms of
four independent structure functions. There are two possible choices of these
functions~\cite{LamTung1}. Lorentz-invariant structure functions are basic ones. For our purpose,
more convenient are so-called helicity structure functions $W_\mathrm{L}, W_\mathrm{T}, W_\mathrm{TT}, W_\mathrm{LT}$. In
this approach, one contracts both hadronic and leptonic tensors with virtual photon
polarization vectors (PPVs). Then, the leptonic degrees of freedom reduce to an angles
${\mit\Omega}=(\theta,\phi)$ in lepton pair center-of-mass frame. $W$ structure functions
correspond to the hadronic degrees of freedom.

The inclusive DY cross section is given by the formula
\begin{eqnarray}
\frac{d\sigma}{d x_\mathrm{F} dM^2 d {\mit\Omega} d^2 q_\bot}&=& \frac{\alpha^2_\mathrm{em}\sigma_0}{2(2\pi)^4
M^4} \Big[ W_\mathrm{L} \left(1-\cos ^2 \theta\right) +   W_\mathrm{T} \left(1+\cos ^2 \theta\right) \nonumber  \\
&&+ W_\mathrm{TT}\left(\sin^2\theta \cos 2\phi\right)+W_\mathrm{LT}\left(\sin2\theta \cos \phi\right) \Big]\,,
\label{sigAsWcomb}
\end{eqnarray}
where $x_\mathrm{F}$ is a fraction of projectile's longitudinal momentum taken by virtual photon,
$M$ is an invariant mass of leptons pair, $q_\bot$ is transverse momentum of virtual
photon and the constant $\sigma_0$ gives the dimension.

Helicity structure functions, in contrast to invariant ones, depend on the choice of axes
in lepton pair center-of-mass frame. Here, the most convenient one is the so-called
$t$-channel helicity frame (see~\cite{LamTung1} for the detailed definition).

\vspace{-2mm}
\section{Inclusive cross section for forward Drell--Yan}

\begin{figure}[htb]
\centerline{%
\includegraphics[width=5cm]{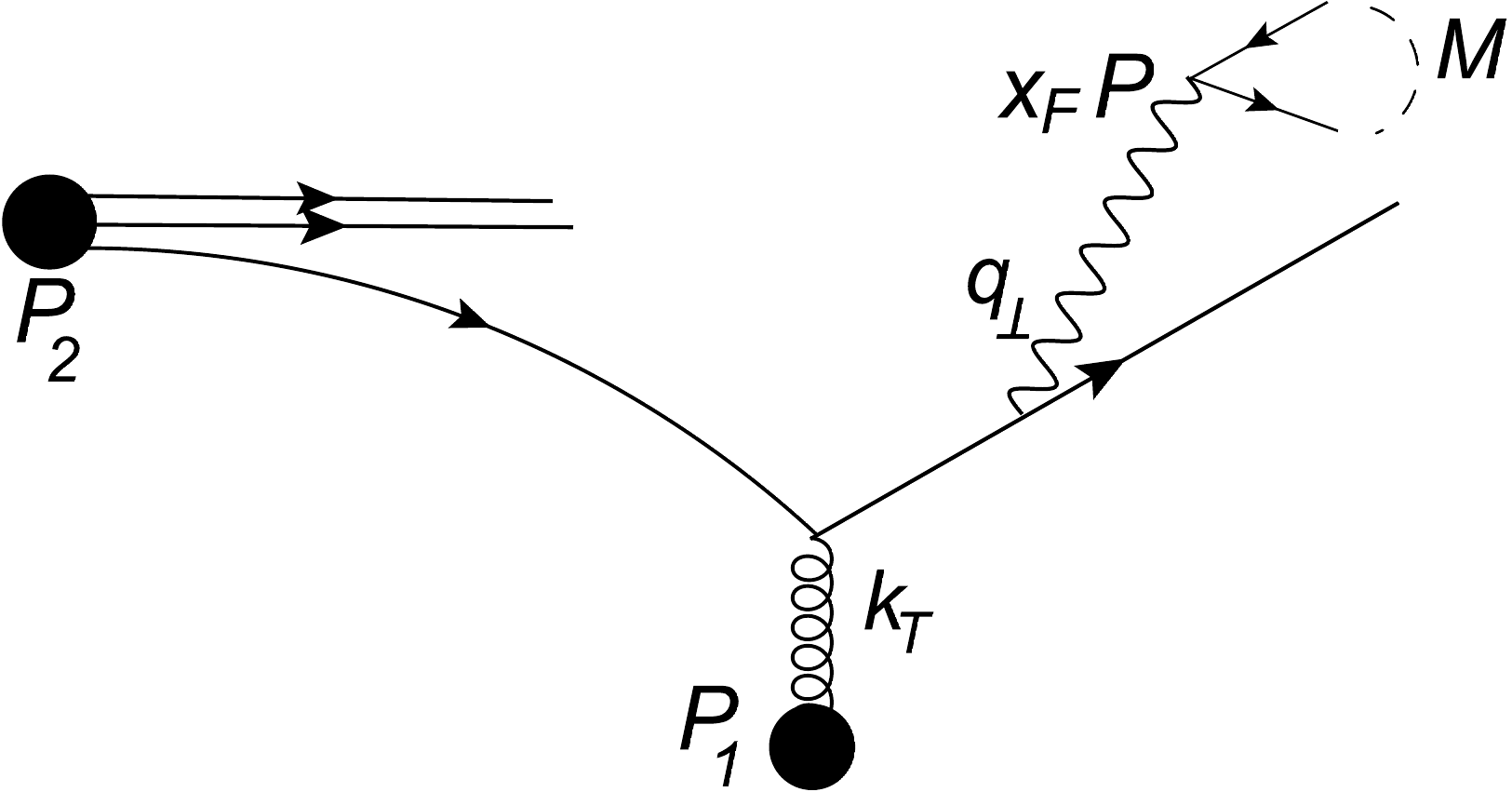}\hspace{5mm}
\includegraphics[width=5.3cm]{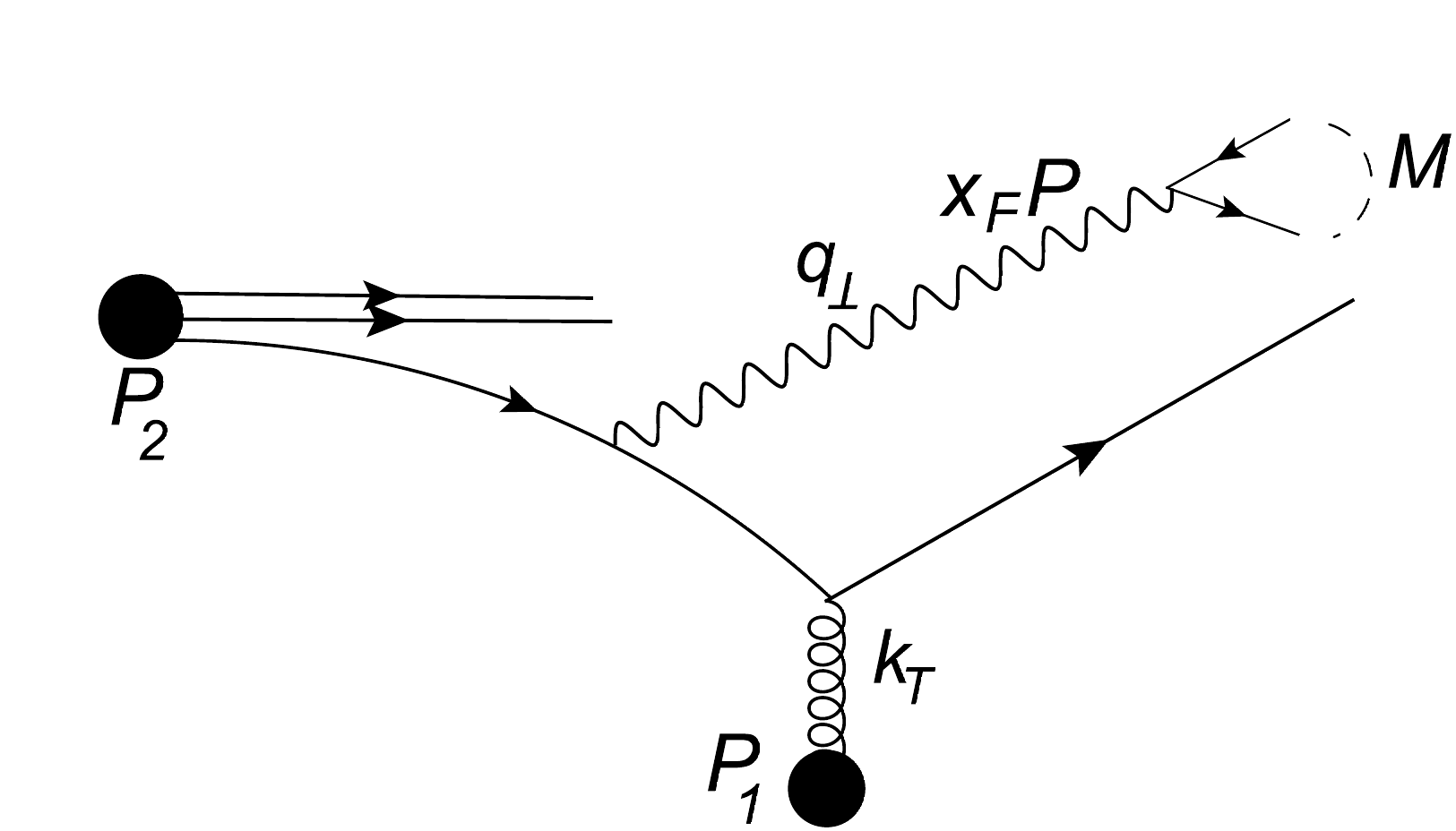}}
\caption{Dominant diagrams for forward Drell--Yan process in target rest frame. Photon
carries $x_\mathrm{F}$ part of projectile's momentum. $M^2$ is photon's virtuality and the square
of invariant mass of lepton--antilepton pair. Transverse momenta of photon and gluon are
denoted as $q_\bot$ and $k_\mathrm{T}$, respectively.}
\label{leadingDiagrams}
\end{figure}

In Fig.~\ref{leadingDiagrams}, we plot leading diagrams for the forward Drell--Yan
scattering in hadron--hadron collisions. They are dominant due to the large gluon density
at small longitudinal momentum fraction of one hadron ($x_1 <10^{-5}$).
The most convenient frame for calculating these diagrams is such that target (proton with
small $x$) is at rest. In this frame, the energy of projectile, $E$, is much larger than
other scales (such as $M$ or $q_\bot$) and we can drop in calculations non-leading terms
in $1/E$ expansion.

To calculate the diagrams from Fig.~\ref{leadingDiagrams}, we apply $k_\mathrm{T}$-factorization
framework, then the differential cross-section can be written it the following way:
\begin{eqnarray}
\hspace{-3mm}\frac{d\sigma}{d x_\mathrm{F} dM^2 d {\mit\Omega} d^2 q_\bot}&=& \frac{\alpha_\mathrm{em} \alpha_\mathrm{s}}{6\pi(P_1\,
P_2)^2 \ M^2}  \int\limits_{x_\mathrm{F}}^1 dz \frac{1}{1-z} \frac{\wp(x_\mathrm{F}/z)}{x_\mathrm{F}^2}  \nonumber \\
\hspace{-3mm}&&\times
\int d^2 k_\mathrm{T} /k_\mathrm{T}^4 \  f\left(\bar{x}_g,k_\mathrm{T}^2\right)  L^{\tau \tau'}({\mit\Omega}) \tilde{\mit\Phi}_{\tau
\tau'} (q_\bot,k_\mathrm{T},z)\,,
\label{dsigma}
\end{eqnarray}
where  $\wp(x_\mathrm{F}/z)$ is a pdf for projectile $P_2$, $f(\bar{x}_g,k_\mathrm{T}^2)$  is an
unintegrated gluon density of target $P_1$, $ L^{\tau \tau'}({\mit\Omega})$ is a lepton tensor
contracted with PPV which reduces to angular coefficients from (\ref{sigAsWcomb}), the
impact factor $\tilde{\mit\Phi}_{\tau \tau'} (q_\bot,k_\mathrm{T},z)$ is a square of hard amplitudes of
diagrams (describing emission of virtual photon).

For our purpose, it is convenient to use the color dipole model in which the unitegrated
gluon density  $f(\bar{x}_g,k_\mathrm{T}^2)$ in (\ref{dsigma}) is replaced by an (equivalent in the
leading logarithmic approximation) color dipole cross section $\hat{\sigma}(r)$~\cite{NZ,Brodsky,Kopeliovich}
\begin{equation}
\int d^2 k_\mathrm{T} /k_\mathrm{T}^4 \  f\left(\bar{x}_g,k_\mathrm{T}^2\right) \  \tilde{\mit\Phi} \left(q_\bot,k_\mathrm{T},z\right)
\rightarrow       \int d^2 r \ \hat{\sigma}(r) {\mit\Phi} (q_\bot,r,z)\,.
\end{equation}

To perform twist expansion of helicity structure functions, we follow methods developed in
Refs.~\cite{twist1,twist2,GolecLew} and apply the Mellin transformation. Then,
(\ref{sigAsWcomb}) is given by
\begin{equation}
W_{i}=\int\limits_{x_\mathrm{F}}^1 dz \ \wp(x_\mathrm{F}/z)
\int\limits_C \frac{ds}{2\pi i} \ \left ( \frac{z^2 Q_0^2}{M^2 (1-z)} \right) ^s \tilde{\sigma}
(-s) \hat{\mit\Phi}_{i} (q_\bot,s,z)\,,
\label{Wphihat}
\end{equation}
where $\tilde{\sigma}(-s)$ and $\hat{\mit\Phi}_{i} (q_\bot,s,z)$ are Mellin transforms of
dipole cross section and impact factor. For the details, see~\cite{MSS}, in particular
expressions for  $\hat{\mit\Phi}_{i} (q_\bot,s,z)$.

\section{Twist expansion}

In formula~(\ref{Wphihat}), we have explicitly three energy scales: saturation scale
$Q_0$ (coming from dipole cross section $\hat{\sigma}$) which is a soft scale and two
semi-hard scales: $M$ and $q_\bot$. OPE is here given in terms of positive powers of the
$Q_0$.

In order to perform twist expansion, one should choose a model of the dipole cross section
$\hat{\sigma}$. We adopt the Golec-Biernat and W\"usthoff model~\cite{GolecWusthoff}
which was proven to be successful in description of Deep Inelastic Scattering (DIS) and
diffractive DIS data from HERA.

Mellin transform of such function is particularly simple $\tilde{\sigma}(\!-s)\!=\! -\sigma_0
{\mit\Gamma}(\!-s)$ and integral~(\ref{Wphihat}) is a sum of infinite number of residues which
are proportional to $Q_0^{2k}$ with $k=1,2,\ldots$\ \,Expressions for twist two $W_i^{(2)}$
and twist four $W_i^{(4)}$ for all structure functions are given in~\cite{MSS}.

To compare the twists, it is useful to introduce averaged cross section over the angles (at
the leading twist)
\begin{equation}
\left<\sigma \right>=\frac{1}{4 \pi} \int  d{\mit\Omega} \ \frac{d\sigma}{d x_\mathrm{F} dM^2 d {\mit\Omega}
d^2 q_\bot} \equiv \frac{2}{3} \left( W_\mathrm{L}^{(2)}+2 W_\mathrm{T}^{(2)} \right)\,.
\label{sigAve}
\end{equation}

\begin{figure}[htb]
\centerline{%
\includegraphics[width=6cm]{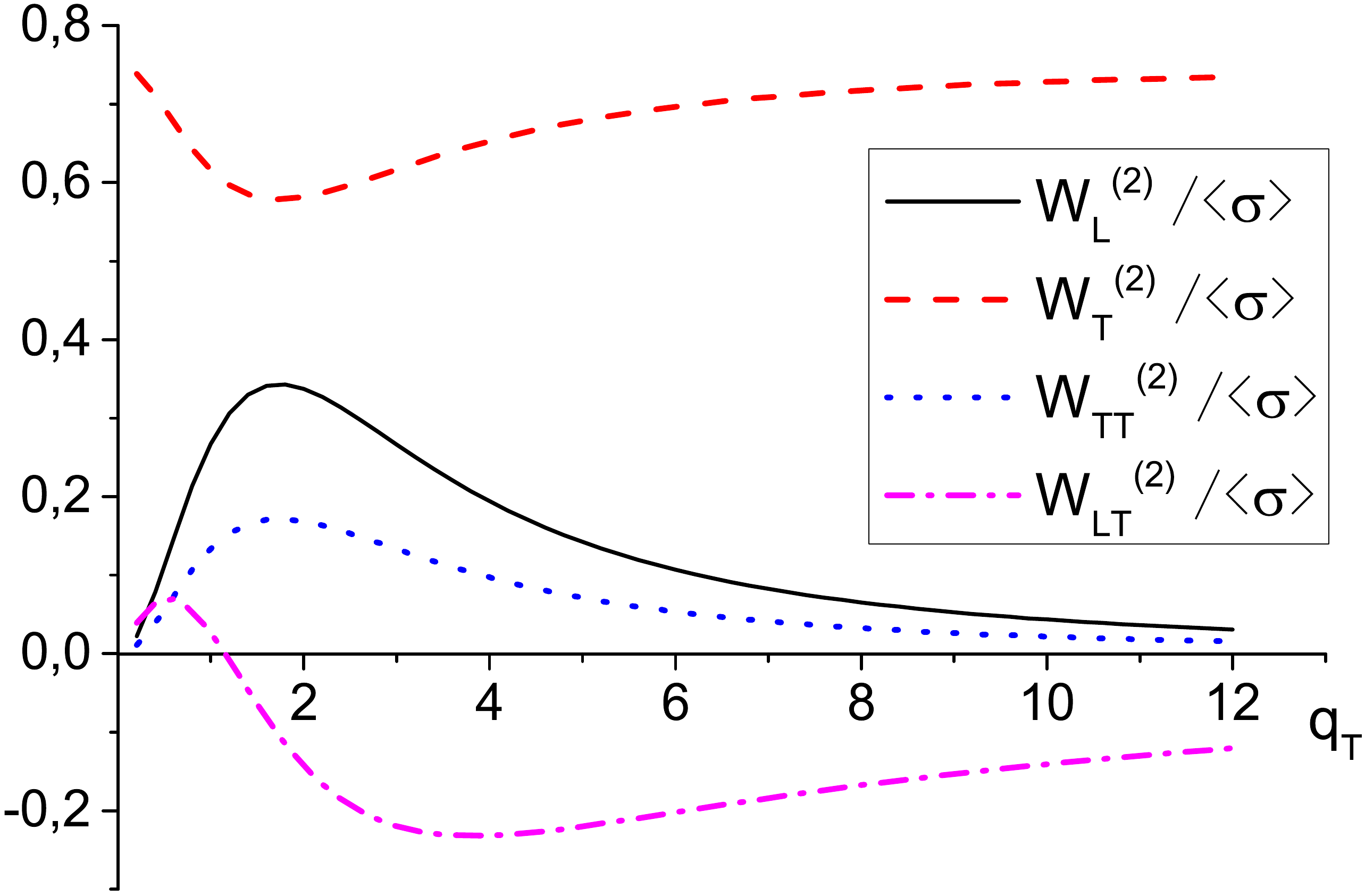}\hspace{5mm}
\includegraphics[width=6cm]{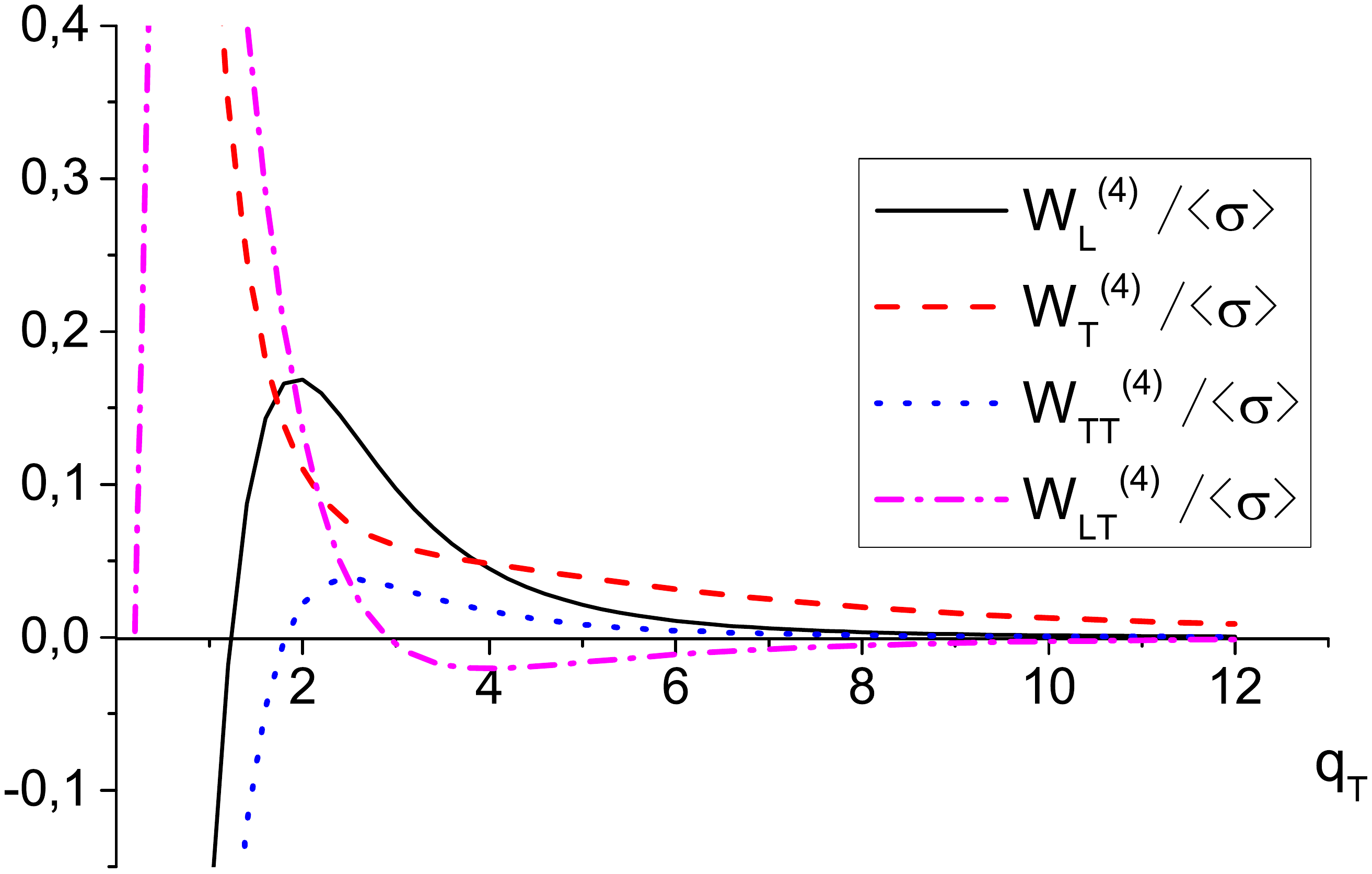}}
\caption{Left: twist 2 divided by averaged cross section $\left<\sigma \right>$ as a
function of photon transverse momentum $q_\mathrm{T}$ for all helicity structure functions. Right:
the same for twist 4. Both plots for $M^2=6$ GeV$^2$ and $x_\mathrm{F}=0.1$.}
\label{tw2i4doAv}
\end{figure}

In Fig.~\ref{tw2i4doAv}, we show twists 2 (left) and 4 (right) divided by $\left<\sigma
\right>$. They are plotted as functions of lepton-pair transverse momentum $q_\mathrm{T}$ for fixed
pair mass $M^2=6$ GeV$^2$, which might be reachable at the LHCb experiment~\cite{lhcb0}.
It can be seen that the $W_\mathrm{T}$ is a dominant structure function at twist 2. The same is at
twist 4 for large $q_\mathrm{T}$, however, for small transverse momentum $W_\mathrm{L}^{(4)}$ is larger.
One should note that around $q_\mathrm{T}\approx 2$ GeV twists 4 start to be very large. This is,
however, a region where $q_\mathrm{T}\approx Q_0$ and twist expansion breaks.

\begin{figure}[htb]
\centerline{%
\includegraphics[width=7cm]{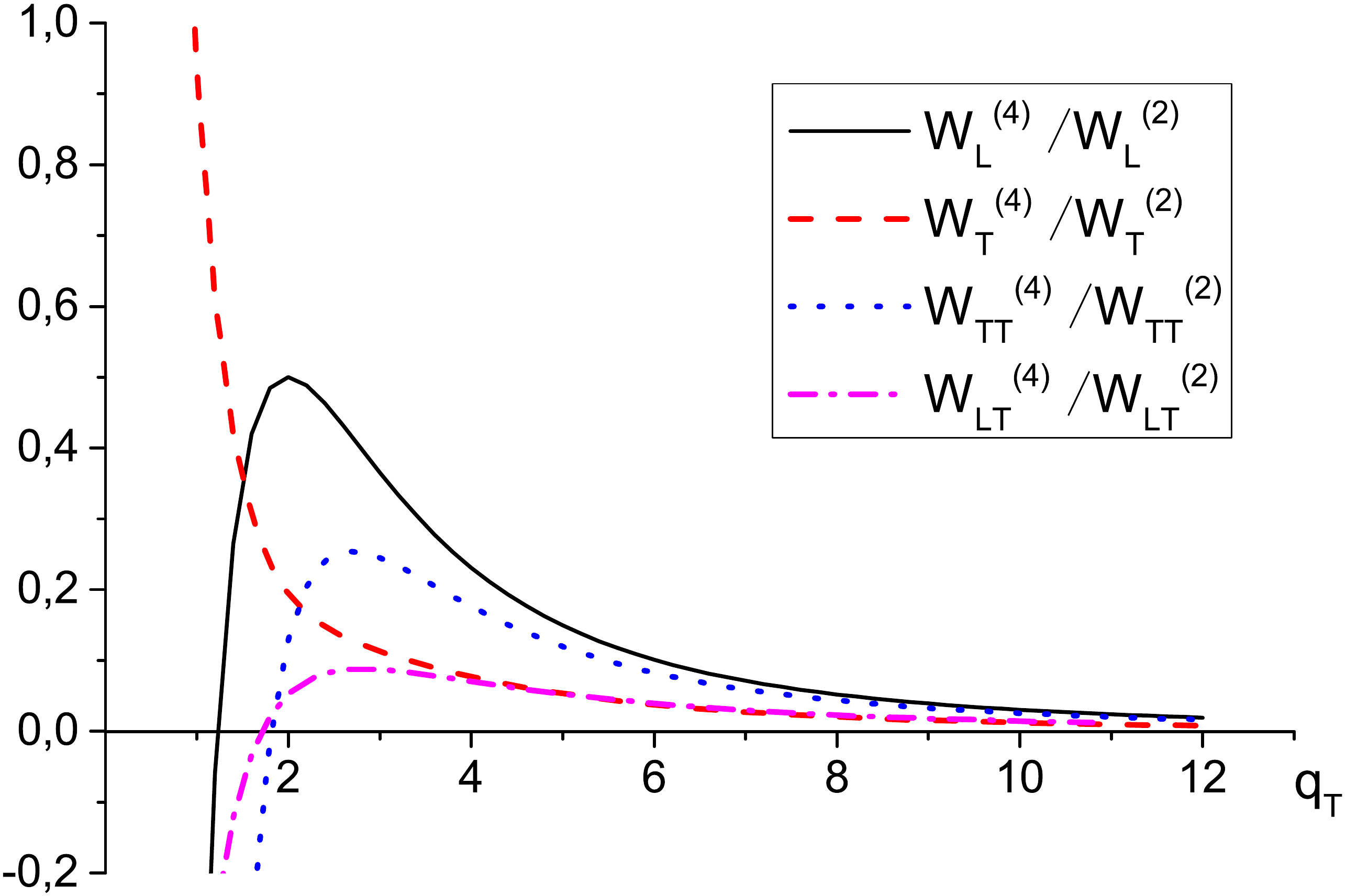}}
\caption{Twist 4 divided by twist 2 as a functions of photon transverse momentum $q_\mathrm{T}$
for all helicity structure functions. $M^2=6$ GeV$^2$ and $x_\mathrm{F}=0.1$.}
\label{tw4dotw2}
\end{figure}

In  Fig.~\ref{tw4dotw2}, we plot ratios of twist 4 to twist 2. It can be seen that the
largest contribution to non-leading twist is for $W_\mathrm{L}$ --- up to 20\% for small~$q_\mathrm{T}$.

\vspace{2mm}

\newpage
\section{Lam--Tung relation}

For experimental searches of higher twists, the most interesting are quantities that vanish
at the leading twist and are nonzero at higher twists.
In the DY process, such a quantity might be constructed using the Lam--Tung
relation~\cite{LamTung2,Gelis}
\begin{equation}
{\mit\Delta}_\mathrm{LTT}^{(2)} \equiv W_\mathrm{L}^{(2)}-2 W_\mathrm{TT}^{(2)}=0\,.
\label{relLT}
\end{equation}
At the next-to-leading twist, namely twist 4, relation is broken (see~\cite{MSS}). It is
also violated by higher order QCD corrections, however, at the very small~$x$, the
contribution coming from higher twists is sizable comparing to them.

\begin{figure}[htb]
\centerline{%
\includegraphics[width=7cm]{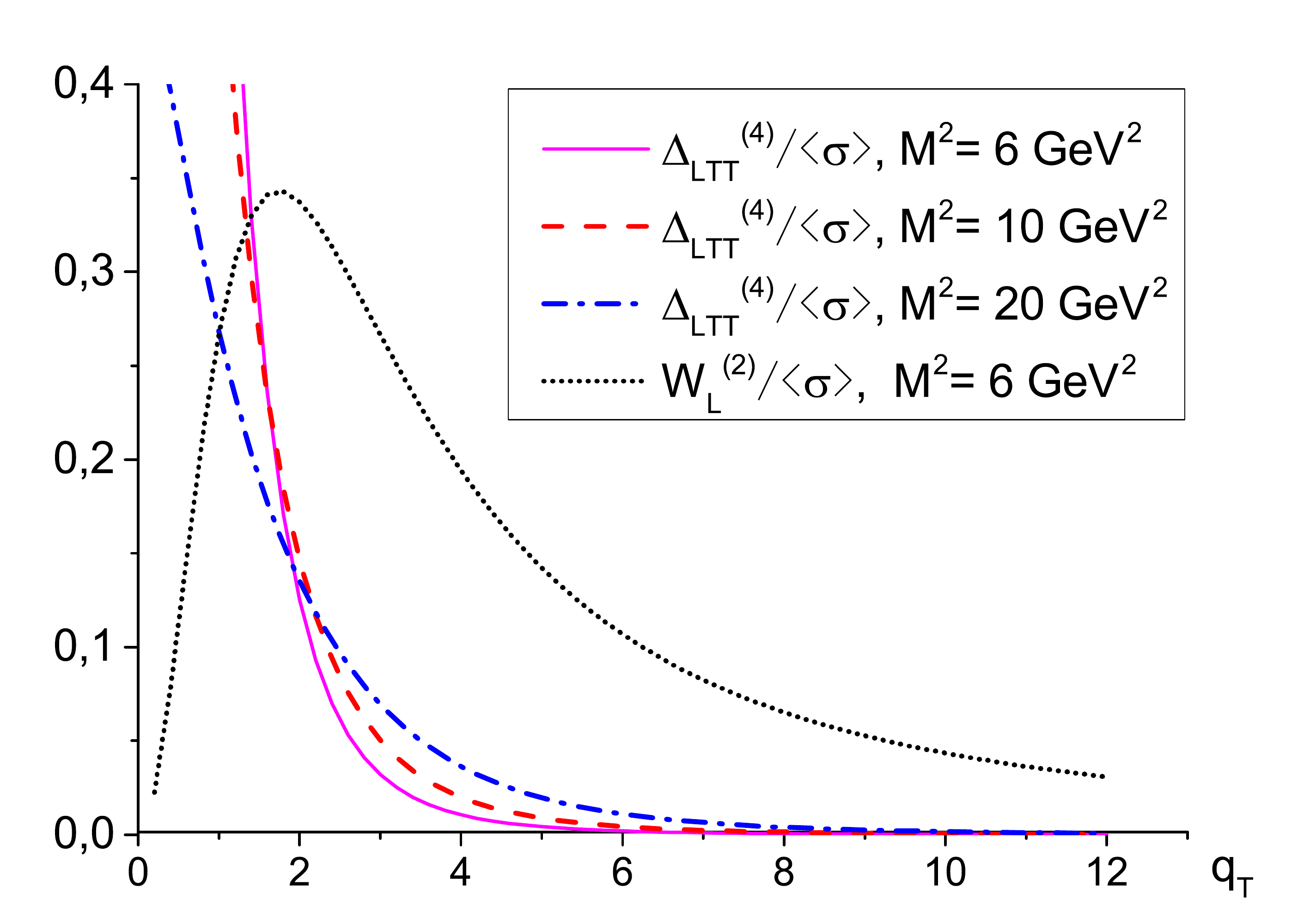}}
\caption{Ratio ${\mit\Delta}_\mathrm{LTT}^{(4)}/ \left<\sigma \right>$ as a function of $q_\mathrm{T}$ for three
masses $M^2$. For comparison: leading twist of $W_\mathrm{L}$ divided by $\left<\sigma \right>$.}
\label{LamTung}
\end{figure}

In Fig.~\ref{LamTung}, we plot ${\mit\Delta}_\mathrm{LTT}^{(4)} \equiv W_\mathrm{L}^{(4)}-2 W_\mathrm{TT}^{(4)}$
divided by $\left<\sigma \right>$ (\ref{sigAve}) as a function of $q_\mathrm{T}$ for different
masses $M^2$. For the comparison, we plot also leading twist of $W_\mathrm{L}$ (black dotted line).
We see that for small transverse momentum, the ratio ${\mit\Delta}_\mathrm{LTT}^{(4)}/W_\mathrm{L}^{(2)}$ could be
around 20\%.

\section{Conclusions and outlook}

The forward Drell--Yan scattering is a promising process for searching of higher twists at
the LHC.
Here, we presented several plots with predictions for higher twists based on the GBW model.
In particular, the quantity $W_\mathrm{L}-2 W_\mathrm{TT}$ should be very useful for searches of higher
twists since it is nonzero only at the next-to-leading twist. For the experimental
searches, it is essential to measure precise angular distribution at low mass and
transverse momentum of lepton pair. Then, as we showed, non-leading twists contribution
could be around 20\%.

\newpage
Authors would like to thank the organizers of the EDS Blois 2015, Corsica, France, for the
very interesting meeting.
Support of the Polish National Science Centre grant No.~DEC-2014/13/B/ST2/02486 is
gratefully acknowledged.
T.S.~acknowledges the support from the scholarship of Marian Smoluchowski Scientific Consortium
``Matter-Energy-Future'' from the KNOW funding.

\bibliographystyle{aipproc} 

\begin{thebibliography}{9}

\bibitem{lhcb0}
  J.~Anderson [LHCb Collaboration],
Proceedings of 40th International Symposium on Multiparticle Dynamics (ISMD 2010) 	21-25 Sep 2010. Antwerp, Belgium.


\bibitem{LamTung1}
C.~S. Lam, and Wu-Ki Tung, \emph{Phys. Rev. D} \textbf{18} (1978) 2447.

\bibitem{NZ} N. N. Nikolaev and B. G. Zakharov, \emph{Z. Phys.} \textbf{C49} (1991) 607.



\bibitem{Brodsky}
S.~J. Brodsky, A. Hebecker, E. Quack, \emph{Phys. Rev. D} \textbf{55} (1997) 2584-2590.


\bibitem{Kopeliovich}
  B.~Z.~Kopeliovich, J.~Raufeisen and A.~V.~Tarasov,
  \emph{Phys.\ Lett.\ B} {\bf 503} (2001) 91.



\bibitem{twist1}
  J.~Bartels, K.~J.~Golec-Biernat and K.~Peters,
  \emph{Eur.\ Phys.\ J.\ C} {\bf 17} (2000) 121.


\bibitem{twist2}
  J.~Bartels, K.~Golec-Biernat and L.~Motyka,
  \emph{Phys.\ Rev.\ D} {\bf 81} (2010) 054017.

\bibitem{GolecLew}
K.~J. Golec-Biernat, E. Lewandowska, A.~M. Stasto, \emph{Phys. Rev. D} \textbf{82} (2010) 094010.

\bibitem{MSS} 
L.~Motyka, M.~Sadzikowski and T.~Stebel,  \emph{JHEP} \textbf{05} (2015) 087

\bibitem{GolecWusthoff}
K.~J. Golec-Biernat, M. W{\"u}sthof, \emph{Phys. Rev. D} \textbf{59} (1998) 014017;
  \emph{Phys. Rev. D} {\bf 60} (1999) 114023.


\bibitem{LamTung2}
C.~S. Lam, and Wu-Ki Tung, \emph{Phys. Lett. B} \textbf{80} (1980) 228.

\bibitem{Gelis}
F. Gelis, J. Jalilian-Marian, \emph{Phys. Rev. D} \textbf{76} (2007) 074015. 

\end{thebibliography}

\end{document}